\documentclass[10pt]{article}
\usepackage{makeidx}
\usepackage{multirow}
\usepackage{multicol}
\usepackage[dvipsnames,svgnames,table]{xcolor}
\usepackage{graphicx}
\usepackage{ulem}
\usepackage{setspace}
\usepackage{amsmath}
\usepackage{amssymb}

\title{}
\usepackage[paperwidth=612pt,paperheight=792pt,top=72pt,right=72pt,bottom=72pt,left=72pt]{geometry}

\makeatletter
	{\par\setlength{\parindent}{#3}
	\setlength{\leftmargin}{#1}       \setlength{\rightmargin}{#2}%
	\advance\linewidth -\leftmargin       \advance\linewidth -\rightmargin%
	\advance\@totalleftmargin\leftmargin  \@setpar{{\@@par}}%
	\parshape 1\@totalleftmargin \linewidth\ignorespaces}{\par}%
\makeatother

\begin{document}

\title{Scale-free interpersonal influences on opinions in complex systems}
\author{Noah E. Friedkin \textsuperscript{1}} 
\maketitle

\noindent \textit{ \textsuperscript{1} Center for Control, Dynamical Systems and Computation, College of Engineering, and  Department of Sociology, College of Letters and Science, University of California, Santa Barbara  }.

\begin{abstract}
\noindent  An important side effect of the evolution of the human brain is an increased capacity to form opinions in a very large domain of issues, which become points of aggressive interpersonal disputes. Remarkably, such disputes are often no less vigorous on small differences of opinion than large differences. Opinion differences that may be measured on the real number line may not directly correspond to the subjective importance of an issue and extent of resistance to opinion change. This is a hard problem for field of opinion dynamics, a field that has become increasingly prominent as it has attracted more contributions to it from investigators in the natural and engineering sciences. The paper contributes a scale-free approach to assessing the extents to which individuals, with unknown heterogeneous resistances to influence, have been influenced by the opinions of others.       
\end{abstract}

\section{Introduction}

Observations of opinion changes on an issue are a necessary basis of asserting the existence of some coordinative mechanism of interpersonal influence in which individuals' opinions are affecting other individuals' opinions. The classical literature on interpersonal influence was developed with experimental designs with subjects exposed to a fixed displayed opinion of one or more confederates of the experimenter \cite{Hovland1953,Asch1956,Milgram2009}.  In such designs, where exogenous disturbances on opinions are minimal, 
interpersonal influence is apparent when subjects' displayed opinions are altered in the direction of the displayed fixed position to which each subject has been exposed. 

However, observed extents of opinion differences on the real number line, and opinion changes, depend the scaling of the opinions and the times at which they are measured,

\begin{equation}
{\Delta (ij,t)} =  {\alpha} \big[  x(i,t+k) - x(j,t) \big], \; j \in \{i,j\}, \; t = 0,1,2,... \; ,
\end{equation}

\noindent where $\alpha > 0$ is the scalar value. As $t$ and/or $k$ increase, $\Delta (ii,t)$ typically decreases as in Sherif's  \cite{Sherif1936} seminal investigation of norm formation, and the observed amount of opinion change will depend on $\alpha$ for all $\Delta (ii,t) > 0$. The sensitivity of $\Delta (ii,t)$ to arbitrary selection of $ \{ t,k \} $ temporal measurement points is ameliorated when $ {\bf{x}}(t) = {\bf{x}}(0)$ are individuals' pre-process initial opinions  and $\mathop {\lim }\limits_{k \to \infty } {{\bf{x}}(t+k)}= {\bf{x}}(\infty)$ are their post-process equilibrium opinions. It is aggravated when it is assumed that interpersonal influence declines as the absolute value of $\Delta (ij,t)$ increases, or that interpersonal influence does not exist above threshold values of $ \mid \Delta (ij,t) \mid$, especially when these thresholds may vary among individuals \cite{Hegselmann2002,Deffuant2000,Lorenz2007}. Here, I shall assume that individuals' quantitative differences of opinion are not informative of their levels of resistance to opinion change. 

A markedly enlarged scientific community is now engaged in the study of complex systems of interpersonal influences composed of individuals with heterogeneous initial opinions (some or all of whom are responding to the changing opinions of others) and heterogeneous extents of resistance to interpersonal influence. With the influx of investigators from the natural and engineering sciences into the field of opinion dynamics, the
development of mathematical models of opinion dynamics is a rapidly advancing frontier of scientific work \cite{Jackson2008,Acemoglu2013,Bindel2011}.
The collection of observations, which was limited to
small groups, now includes the behaviors and networks of communicating individuals in large-scale field settings of Internet social media. The present results contribute an approach to a scale-free derivation of individuals' heterogeneous resistances to interpersonal influences in complex systems. 

Finite convex sets are important in various areas of basic and applied mathematics, and appear in linear state-space processes. One area of application is the first-order DeGroot \cite{DeGroot1974} discrete-time state-space process in which the state of each point of a set $n$ points, ${\bf{x}}_{i}(k+1),\; i = 1,...,n$, is a convex combination of the immediately prior states of all points of the system ${\bf{x}}_{j}(k)  \in  \mathbb{R}^{n},\; j = 1,...,n$. With a $x(0) \in \mathbb{R}^n$ and a row-stochastic ${\bf{W}}$, the model

\begin{equation}
{\bf{x}}(t+1) = {\bf{W}}{\bf{x}}(t), \;\; t=0,1,... \; ,
\end{equation}

\noindent was formulated as a  mechanism by which consensus might be reached among a set of individuals. It has become the benchmark model of the literature on opinion dynamics. Its precursors include the models of French \cite{French1956} and Harary \cite{Harary1959}. In addition, the model has become increasingly prominent in control theory \cite{Bullo2009,Huang2009,Moreau2005,Olfati2007}. When the mechanism unfolds in an aperiodic irreducible network, the system converges to a single value on the real number line and, more generally for a ${\bf{X}}(k) \in \mathbb{R}^{n \times m}$, to a single location in $m$-dimensional space. 

It will be shown that scale-free interpersonal influences are detectable for this model. Note that the DeGroot model assumes measures of ${\bf{x}}(0)$ and ${\bf{W}}$. In it, the main-diagonal values of ${\bf{W}}$ correspond to individuals' resistance to interpersonal influence. In practice, a direct measure of these main-diagonal values are rarely available, so that the measure of ${\bf{W}}$ is a matrix with (a) zeros on its main-diagonal and (b) off-diagonal values that are relative \textit{interpersonal} weights based on the available network data. Thus, the usual implementation of the model assumes individuals with zero resistance, who adopt the weighted average of \textit{others'} positions on an issue at each time $t$. But if the system is, in truth, composed of individuals with high resistances to opinion change, then the process of consensus formation may be exceedingly slow and difficult. With high resistances to opinion change, the process may be terminated prior to its convergence to consensus. 

Hence, the analytical problem may formalized as follows. Let ${\bf{C}}$ be a row stochastic matrix of relative \textit{interpersonal} weights with a zero main diagonal. Given a measure of such a matrix and measures of individuals' opinions, what are their' resistances to opinion change?     

\section{Scale-free Resistance in the DeGroot Model}

With little loss of generality, let $0 < {w_{ii}}<1$ for all $i=1,...,n$. Let ${\bf{D}} \equiv \text{diag}({w_{11}},...,{w_{nn}})$.  Let ${\bf{C}} \equiv ({\bf{I}}-{\bf{D}})^{-1} ({\bf{W}}- {\bf{D}})$. Hence, ${\bf{C}}= [{c_{ij}}] $ is a matrix with ${c_{ii}}=0$ for all $i$ and ${c_{ij}}={w_{ij}} / (1-{w_{ii}}) $ for all $i \ne j$. N.B. The foregoing are the theoretical definitions of constructs. The matrix ${\bf{D}}$ is the unknown values of the main-diagonal of ${\bf{W}}$, and the matrix ${\bf{C}}$ is the available measure of the relative interpersonal influences of ${\bf{W}}$. 

From the definitions of these constructs, it follows that 

\begin{equation}
{\bf{W}} = ({\bf{I}} - {\bf{D}}){\bf{C}} + {\bf{D}} 
\end{equation}
 
\noindent with which a solution of the resistance values of ${\bf{D}}$ is available.

The scalar equation of the DeGroot model may now be expressed in terms of ${\bf{C}}$ and ${\bf{D}}$

\begin{align}
{x_{i}(t + 1)} & =
{\sum\limits_{j = 1}^n} {{w_{ij}}{x_{j}(t)}},  
\, \, \, i = 1,2,...,n, \; t = 0,1,2,...\, \, . \notag \\
& = ({d_{ii}}){x_{i}}(t) + ( 1- {d_{ii}})\sum\limits_{j \ne i} {{c_{ij}}{x_j}(t)} , 
\end{align}

\noindent whence

\begin{equation}
{d_{ii}} = 1 -  {\frac{ {x_{i}}(t+1) - {x_{i}}(t) } { \sum\limits_{j \ne i} {{c_{ij}}{x_j}(t)}  - {x_{i}}(t)} } ,  
\end{equation}                      
  
\noindent for ${ \sum\limits_{j \ne i} {{c_{ij}}{x_j}(t)}  \ne {x_{i}}(t)} $. 

Under the assumption of a time-invariant ${\bf{W}}$, the derived ${d_{ii}}$ values, $i = 1,2,...,n$, apply to all $t$. This assumption may be relaxed with a ${\bf{W}}(t)$. In either case, the derived values are scale-free. In applications, solutions are constrained to the $ ( 0,1 ) $ interval. Under the assumption that the model is correct, unreliable or invalid measurement models of its constructs may generate departures from the $ ( 0,1 ) $ interval, and unreliable or invalid ${d_{ii}}$ values.

\section{Discussion}

A model-based approach has been presented for a sale-free  detection of extents to which individuals' opinions are closed or open to interpersonal influences on their issue-specific opinions. 
Influence detection does not require full data on the network
in which opinion dynamics unfold. Because only direct influences on an individual are involved in the detection, the approach may be applied to a selected set (or random sample) of individuals and their adjacent neighbors. 

The class of models considered in this paper is based on the assumption that endogenous interpersonal influence is a convex combination mechanism with weights that are allocated by individuals to their own and others' displayed orientations to an object. The available empirical supports for this mechanism are inconsistent with an assumption that allows negative interpersonal influences or resistances to influences. The mechanism predicts (a) an initial consensus will be maintained and (b) that modifications of disagreeing initial opinions will be constrained to range (or convex hull) of initial opinions. These predictions are strongly supported in the observations of experimental social psychology collected on small groups assembled in laboratory settings, where individuals' pre- and post-discussion opinions have been measured.  

(a) There is strong support in the literature for the  prediction that an initial consensus will be maintained. Barnlund \cite{Barnlund1959}  reported that, in small groups assembled to solve problems of logic, an initial consensus was not questioned (the group moved on to the next problem) regardless of whether the consensus was correct or incorrect. Similarly, Thorndike \cite{Thorndike1938} found that an initial consensus was rarely modified regardless of whether the consensus was correct or incorrect; in his results, an initial consensus was modified in only 3 of 725 group problem-solving trials in which the group’s judgment was correct, and in 1 of 263 trials in which the group’s judgment was incorrect. Consensus is either assumed to be correct (whether or not it is) or satisfactory; in either case, it is deemed conclusive.

(b) Given initial disagreement, Friedkin and Johnsen \cite{FJ2011} find one instance of an individual post-discussion opinion outside the range of initial opinions in 1,000 cases based on 50 groups of tetrads discussing five issues (two monetary issues and three issues of acceptable risk) in sequence. Similarly, they find one instance of an individual post-discussion opinion outside the range of initial opinions in 288
cases based on 32 groups of triads discussing three issues of acceptable risk in sequence.  A larger incidence of anomaly occurs in dyads, and appears to be linked to the special property of this smallest of groups in which an aperiodic
influence system is approached as their members' allocations of weight to their own opinions approach 0. 

An additional important property of the class of models considered in this paper is that interpersonal influence is independent of $ \mid \Delta (ij,t) \mid$. The available evidence on this matter, amassed over decades, is mixed \cite{FJ2011} and erodes the assertion of a reliable dependency. The subjective importance that individuals assign to particular opinion differences $ \mid \Delta (ij,t) \mid$ appears to be unconstrained by the values of these differences, so that large resistances to opinion change may exist for small differences of opinions, and small resistances to opinion change may exist for large differences of opinions.     

From a social psychological perspective, convex combination mechanisms describe an automatic heuristic ``cognitive algebra'' \cite{Anderson1974, Anderson1991} with which the brain integrates heterogeneous information. Thus, the influence network is a social cognition structure assembled by individuals' allocation of weights to self and particular others on a specific issue. Investigators in the discipline of experimental social psychology have been remarkably adept in their detection numerous conditions with effects on the interpersonal influence relation. The implication of this work is that the antecedents of individuals'  allocated weights are complex, i.e., the weights of the convex combination mechanism are net resultants of numerous  conditions of the individuals and their interpersonal relations. As such, the set of ordered pairs of weights, which form an influence network, is unlikely to be based on any single condition, e.g., opinion differences, expertise, authority, rewards, affection, or coercion.

\bibliography{ArticleReferences}

\begin{thebibliography}{10}

\bibitem{Hovland1953}
Carl~I Hovland, Irving~L Janis, and Harold~H Kelley.
\newblock {\em Communication and persuasion: psychological studies of opinion
  change.}
\newblock Yale University Press, 1953.

\bibitem{Asch1956}
Solomon~E Asch.
\newblock Studies of independence and conformity: I. a minority of one against
  a unanimous majority.
\newblock {\em Psychological Monographs: General and Applied}, 70(9):1--70,
  1956.

\bibitem{Milgram2009}
Stanley Milgram.
\newblock {\em Obedience to authority: An experimental view}.
\newblock Harpercollins, 2009.

\bibitem{Sherif1936}
Muzafer Sherif.
\newblock {\em The psychology of social norms}.
\newblock Harper, 1936.

\bibitem{Hegselmann2002}
Rainer Hegselmann and Ulrich Krause.
\newblock Opinion dynamics and bounded confidence models, analysis, and
  simulation.
\newblock {\em Journal of Artificial Societies and Social Simulation}, 5(3),
  2002.

\bibitem{Deffuant2000}
Guillaume Deffuant, David Neau, Frederic Amblard, and G{\'e}rard Weisbuch.
\newblock Mixing beliefs among interacting agents.
\newblock {\em Advances in Complex Systems}, 3(01n04):87--98, 2000.

\bibitem{Lorenz2007}
Jan Lorenz.
\newblock Continuous opinion dynamics under bounded confidence: A survey.
\newblock {\em International Journal of Modern Physics C}, 18(12):1819--1838,
  2007.

\bibitem{Jackson2008}
Matthew~O. Jackson.
\newblock {\em Social and Economic Networks}.
\newblock Princetion University Press, 2008.

\bibitem{Acemoglu2013}
Daron Acemoglu, Giacomo Como, Fabio Fagnani, and Asuman Ozdaglar.
\newblock Opinion fluctuations and disagreement in social networks.
\newblock {\em Mathematics of Operations Research}, 38(1):1--27, 2013.

\bibitem{Bindel2011}
David Bindel, Jon Kleinberg, and Oren~Shmuel S.
\newblock How bad is forming your own opinion?
\newblock {\em Proc. 52nd IEEE Symposium on Foundations of Computer Science},
  2011.

\bibitem{DeGroot1974}
Morris~H DeGroot.
\newblock Reaching a consensus.
\newblock {\em Journal of the American Statistical Association},
  69(345):118--121, 1974.

\bibitem{French1956}
John~RP French~Jr.
\newblock A formal theory of social power.
\newblock {\em Psychological review}, 63(3):181--194, 1956.

\bibitem{Harary1959}
Frank Harary.
\newblock A criterion for unanimity in french's theory of social power.
\newblock In Dorwin Cartwright, editor, {\em Studies in Social Power}, pages
  168--182. Institute for Social Research, 1959.

\bibitem{Bullo2009}
Francesco Bullo, Jorge Cortes, and Sonia Martinez.
\newblock {\em Distributed control of robotic networks: a mathematical approach
  to motion coordination algorithms}.
\newblock Princeton University Press, 2009.

\bibitem{Huang2009}
Miny1 Huang and Jonathan~H Manton.
\newblock Coordination and consensus of networked agents with noisy
  measurements: Stochastic algorithms and asympototic behavior.
\newblock {\em SIAM J. Control}, 48(1):134--161, 2009.

\bibitem{Moreau2005}
Luc Moreau.
\newblock Stability of multiagent systems with time-dependent communication
  links.
\newblock {\em IEEE Transactions on Automatic Control}, 50(2):169--182, 2005.

\bibitem{Olfati2007}
Reza Olfati-Saber, Alex Fax, and Richard~M Murray.
\newblock Consensus and cooperation in networked multi-agent systems.
\newblock {\em Proceedings of the IEEE}, 95(1):215--233, 2007.

\bibitem{Barnlund1959}
Dean~C Barnlund.
\newblock A comparative study of individual, majority, and group judgment.
\newblock {\em The Journal of Abnormal and Social Psychology}, 58(1):55, 1959.

\bibitem{Thorndike1938}
Robert~L Thorndike.
\newblock The effect of discussion upon the correctness of group decisions,
  when the factor of majority influence is allowed for.
\newblock {\em The Journal of Social Psychology}, 9(3):343--362, 1938.

\bibitem{FJ2011}
Noah~E Friedkin and Eugene~C Johnsen.
\newblock {\em Social influence network theory: a sociological examination of
  small group dynamics}, volume~33.
\newblock Cambridge University Press, 2011.

\bibitem{Anderson1974}
Norman~H Anderson.
\newblock Information integration theory: A brief survey.
\newblock {\em Contemporary developments in mathematical psychology},
  2:236--305, 1974.

\bibitem{Anderson1991}
Norman~H Anderson.
\newblock {\em Contributions to Information Integration Theory: Cognition},
  volume~1.
\newblock Lawrence Erlbaum, 1991.

\end{thebibliography}

\end{document}